\documentclass[manuscript,screen]{acmart}

\AtBeginDocument{%
  \providecommand\BibTeX{{%
    \normalfont B\kern-0.5em{\scshape i\kern-0.25em b}\kern-0.8em\TeX}}}

\setcopyright{acmcopyright}
\copyrightyear{2022}
\acmYear{2022}
\acmJournal{TOCHI}
\acmDOI{10.1145/3534561}

 \usepackage{multirow}
\usepackage{xcolor,colortbl}
 \usepackage{mdframed}
 \mdfdefinestyle{MyFrame}{
    outermargin=3pt,
    innerrightmargin=3pt,
    innerleftmargin=3pt,
    innertopmargin=4pt,
    innerbottommargin=4pt
}
\definecolor{maroon}{cmyk}{0,0.87,0.68,0.32}
\definecolor{lightgreen}{cmyk}{0.39,0.0,0.39,0.00}
\definecolor{aqua}{rgb}{0.0, 1.0, 1.0}

\begin{document}

\title[Advancing Human-AI Complementarity]{Advancing Human-AI Complementarity: The Impact of User Expertise and Algorithmic Tuning on Joint Decision Making}

\author{Kori Inkpen}
\email{kori@microsoft.com}
\affiliation{\institution{Microsoft Research}
\city{Redmond}
  \state{WA}
 \country{USA}
}

\author{Shreya Chappidi}
\affiliation{\institution{University of Virginia}
\country{USA}}

\author{Keri Mallari}
\affiliation{\institution{University of Washington}
\country{USA}}

\author{Besmira Nushi}
\affiliation{\institution{Microsoft Research}
\country{USA}}

\author{Divya Ramesh}
\affiliation{\institution{University of Michigan}
\country{USA}}

\author{Pietro Michelucci}
\affiliation{\institution{Human Computation Institute}
\country{USA}}

\author{Vani Mandava}
\affiliation{\institution{Microsoft Research}
\country{USA}}

\author{Libuše Hannah Vepřek}
\affiliation{\institution{LMU Munich}
\country{Germany}}

\author{Gabrielle Quinn}
\affiliation{\institution{Western Washington University}
\country{USA}}


\renewcommand{\shortauthors}{Inkpen, Chappidi, et al.}

\begin{abstract}
\section{Abstract}

Human-AI collaboration for decision-making strives to achieve team performance that exceeds the performance of humans or AI alone. However, many factors can impact success of Human-AI teams, including a user's domain expertise, mental models of an AI system, trust in recommendations, and more. This paper reports on a study that examines users' interactions with three simulated algorithmic models, all with equivalent accuracy rates but each tuned differently in terms of true positive and true negative rates. Our study examined user performance in a non-trivial blood vessel labeling task where participants indicated whether a given blood vessel was flowing or stalled. Users completed 150 trials across multiple stages, first without an AI and then with recommendations from an AI-Assistant. Although all users had prior experience with the task, their levels of proficiency varied widely. 
 
Our results demonstrated that while recommendations from an AI-Assistant can aid in users' decision making, several underlying factors, including user base expertise and complementary human-AI tuning, significantly impact the overall team performance. First, users' base performance matters, particularly in comparison to the performance level of the AI. Novice users improved, but not to the accuracy level of the AI. Highly proficient users were generally able to discern when they should follow the AI recommendation and typically maintained or improved their performance. Mid-performers, who had a similar level of accuracy to the AI, were most variable in terms of whether the AI recommendations helped or hurt their performance. Second, tuning an AI algorithm to complement users' strengths and weaknesses also significantly impacted users' performance. For example, users in our study were better at detecting flowing blood vessels, so when the AI was tuned to reduce false negatives (at the expense of increasing false positives), users were able to reject those recommendations more easily and improve in accuracy. Finally, users’ perception of the AI's performance relative to their own performance had an impact on whether users' accuracy improved when given recommendations from the AI.  
Overall, this work reveals important insights on the complex interplay of factors influencing Human-AI collaboration and provides recommendations on how to design and tune AI algorithms to complement users in decision-making tasks.

\end{abstract}



\maketitle

\section{Introduction}
Recent advances in artificial intelligence (AI) have inspired promising applications of AI as assistants to people during complex decision-making tasks, including legal, medical, and financial scenarios. In such scenarios, there is an expectation that AI recommendations will help in improving human decisions \cite{10.5555/3061053.3061219, 10.5555/2343576.2343643} and furthermore in forming \emph{complementary} Human-AI teams. In complementary Human-AI teams, it is expected that team performance is better than if the human and the AI each were to operate alone. However, when AI systems and models are trained from historical data, their performance and behavior are rarely optimized or even planned with such collaborative goals in mind and therefore, team performance may be inferior to the performance of the AI alone \cite{bansal2021does, 10.1145/3377325.3377498, Bussone2015TheRO}. For example, if a model is used to give diagnostic recommendations to a medical professional, would it be best if the model is tuned to flag a disease more often than not (even when likelihood is lower) to draw the attention of the human? Or is it best if it only flags a disease when the likelihood is significantly high? Would these behavioral dynamics change if the medical professional had more or less experience in diagnosing the disease? 

Answering these questions prior to deployment or potentially even training of an AI system could help stakeholders choose the right model for deployment—the one that is the best fit for the human collaborator. While for many autonomous applications, the most accurate AI is also the best fit for deployment, this is not always the case for Human-AI collaborative applications. In fact, several studies on Human-AI decision making have presented counterexamples where a less accurate AI may be a better collaborator for the human or when an equally accurate AI may have a variable impact on team performance depending on the predictability of AI mistakes, changes in the AI behavior over time, or the availability and form of algorithmic explanations (e.g., \cite{bansal2019beyond, chiang2021you, gonzalez2020human}). Despite the fact that most models today are optimized with accuracy in mind, these studies indicate that other factors and algorithmic properties may be equally important for Human-AI teams.

In this paper, we set out to expand the understanding of how 
AI recommendations can maximize Human-AI team performance by exploring the impact of (1) \emph{algorithmic tuning} and (2) \emph{human expertise}. We consider algorithmic tuning (i.e., true positive vs. true negative rates) as this is one of the most common control points that AI practitioners have to tune the bias of a trained model. Such adjustments can be implemented in practice either by oversampling positive or negative examples~\cite{han2005borderline}, assigning different costs to different examples during learning~\cite{elkan2001foundations}, or simply by assigning different confidence thresholds for detection depending on whether the goal is to reduce the number of false positives or false negatives (i.e., generally, the higher the confidence threshold, the lower the number of false positives when confidence is reliable). We intersect the study of algorithmic tuning impact with parallel dimensions of human expertise and perceptions of AI to further develop our findings. When the goal is to enable complementary teams, the extent of human expertise and potential tuning that might have happened over time due to application requirements or incentive structures may also impact how human decision-makers make use of an AI assistant. For example, would a human decision maker that has a high (but imperfect) true positive rate (TPR) work best with a high-precision algorithm? Or would they benefit more from a partner with a high (but imperfect) true negative rate (TNR)?

In summary, this paper investigates the following research questions:\\
\noindent {\bfseries RQ1:} What is the impact of tuning the true positive and true negative rates of an AI system on overall Human-AI team performance?
\\
\noindent {\bfseries RQ2:} What is the role of human expertise and predictive tuning in enabling complementary Human-AI teams?
\\
\noindent {\bfseries RQ3:} What strategies do users employ when working with an AI-Assistant in a decision making task, and how do users' perceptions of an AI-Assistant impact Human-AI team performance?

Our study utilized Stall Catchers\footnote{\url{https://stallcatchers.com}}, a citizen-science platform, to explore a complex decision-making task. The goal of Stall Catchers is to produce high-quality label data to accelerate research on Alzheimer’s disease. More specifically, on the platform, citizen science participants are presented with a video clip of blood flow derived from \textit{in vivo} 2-photon excitation microscopy in a mouse brain. The analytic task is to decide whether an indicated blood vessel segment is flowing or stalled~\cite{michelucci2019people, nugent2021accelerating}, and if stalled, indicate the location of the stall in the vessel. 

The Stall Catchers platform was selected for three main reasons. First, the Stall Catchers task is complex for both humans and learned models. This makes the expectation of complementary Human-AI performance more realistic and useful in practice as a future AI system could be deployed to further improve team performance. Second, the task reflects a domain where people have varying degrees of expertise, allowing us to study the influence of varying human expertise. Finally, operating in a citizen-science domain enables us to approximate a real-world decision-making task since the main participant incentive for the project is acceleration of Alzheimer's disease research and participants are aware that their decisions will be used for this purpose. 

Our contributions are as follows:

\begin{itemize}
    \item We demonstrate how users' baseline expertise significantly impacts Human-AI team performance, and that users who demonstrate performance levels similar to that of AI are particularly sensitive to algorithmic tuning.
    
    \item We illustrate how users' perceptions of the AI (relative performance, trust, and usefulness) significantly impact the utility of an AI-Assistant. 

    \item We show that tuning an algorithm's properties (such as adjusting the true positive and true negative rate) must take into account characteristics of the users in order to positively impact Human-AI team performance.
    
    \item We highlight opportunities to develop human-centered algorithms that complement the limitations of human expertise and user perceptions of AI. 

\end{itemize}

This paper is organized as follows. Section~\ref{sec:related_work} situates the work in the context of prior studies and provides background on the Stall Catchers citizen-science project. Section~\ref{sec:study} details the experimental setup along with the different conditions and input data leveraged for the study. Next, Section~\ref{sec:results} presents the study results across the dimensions of different algorithmic tuning and human expertise levels. Parallel to the quantitative analysis, this section also provides a rich set of qualitative insights on how participants perceived and used the AI assistance. Section~\ref{sec:discussion} discusses the presented results, their significance, and known limitations. Finally, we conclude with a set of recommendations and implications for future work in Section~\ref{sec:conclusion}.

\section{Related Work}
\label{sec:related_work}
\subsection{Human AI Collaboration for decision making}
A growing research area in AI involves developing systems that can partner with people to accomplish tasks that exceed the capabilities of the AI alone or the human alone \cite{kamar2016directions, wang2016deep, wang2021deep, steiner2018impact, gaur2016effects}. For example, Steiner et al.~\cite{steiner2018impact} explored the impact of computer assistance in the field of pathology to improve interpretation of images and clinical care \cite{steiner2018impact}. They found that algorithm-assisted pathologists demonstrated higher accuracy than either the algorithm or the pathologist alone and that pathologists considered image review to be significantly easier when interpreted with AI assistance. Wang et al. also demonstrated how deep learning led to a significant improvement in accuracy of pathological diagnoses \cite{wang2016deep}. In a later work, Wang et al. built an AI system that detects pneumonia and COVID-19 severity from chest X-rays, and compared system performance to that of radiologists in routine clinical practice \cite{wang2021deep}. They found that the system helped junior radiologists perform close to the level of a senior, and that average weighted error decreased when the AI system acted as a second reader as opposed to an `arbitrator' between two radiologists.

On the other hand, there is also a set of work where the AI assistance did not lead to any team performance improvements. Lehman et al. measured the diagnostic accuracy of screening mammography with and without a computer-assisted diagnostic (CAD) tool \cite{lehman2015diagnostic}. In their work, they reported sensitivity (true positive) and specificity (true negative) rates before and after the implementation of CAD. They found that CAD was associated with decreased sensitivity and no changes in overall performance or specificity. 

In the following sections, we will highlight prior research exploring the impact of tuning different properties of the AI and human expertise on overall performance of Human-AI teams.

\subsubsection{Findings on the impact of AI and AI properties on team performance} Prior work has demonstrated that overall AI accuracy (both stated and observed) impacts human trust. In this work, Yin et al. found that the effect of stated accuracy, or the model accuracy chosen by researchers, can change depending on the observed model accuracy by participants  ~\cite{yin2019understanding}. Stated accuracy, however, may not always correctly represent an algorithm's performance, especially in real-world deployments with domain shifts. Mismatches between factual and stated algorithmic scores were investigated by De Arteaga et al., who showed that humans are less likely to follow the AI recommendation when the stated score is an incorrect estimate of risk \cite{de2020case}.

AI tuning was later explored by Kocielnik et al. whose work explored two versions of an AI-based scheduling assistant with the same level of 50\% accuracy but with a different emphasis on the types of errors made (avoiding false positives vs. false negatives) \cite{kocielnik2019will}. Interestingly, the study finds that these different modifications can impact humans' perception of accuracy and acceptance of the AI system by exploring techniques for setting expectations including an accuracy indicator, example based explanations, and performance control. However, prior work has not yet investigated the impact of such tuning on the overall team performance or the ability to support complementary teams, which is one of the central research questions we address in this study. 

On a related topic, Bansal et al. \cite{bansal2019beyond,bansal2019updates} show that other AI properties beyond AI accuracy also impact team performance, including the predictability of errors and whether the model errors remain backward compatible over updates. These findings showcase that outside of tuning for true positive or true negative rates, there may exist other types of algorithmic tuning that can better support humans.

\subsubsection{Findings on the impact of human perception on team performance}
A parallel line of work has explored the impact of human perception, such as the role of trust and mental models of the AI, on the overall team performance. 

Dzinolet et al. conducted a series of studies to determine whether people rely on automated and human group members differently \cite{dzindolet2002perceived}. Their self-reported data indicated a bias towards automated aids over human aids, but performance data revealed that humans were more likely to disuse automated aids than to disuse human aids. Schaffer et al. \cite{schaffer2019can} studied the effect of explanations on users' trust in a modified version of the prisoner's dilemma game. They found that explanations not only swayed people who reported very low task familiarity, but showing explanations to people who reported more task familiarity led to automation bias. This finding is further supported by a series of work at the intersection of human-computer interaction and machine learning interpretability, which highlights concerns around the phenomenon of increased, but not appropriate, reliance on automation~\cite{lai2019human, wang2021explanations, bansal2021does, poursabzi2021manipulating, buccinca2021trust}.

Moreover, Zhang et al. \cite{Zhang_2020} discusses trust calibration in AI in detail, as well as the importance of users having a correct mental model of the AI's error boundaries. By definition, the error boundary of a decision-maker (either human or AI) separates the cases for which the decision-maker is correct from the incorrect ones~\cite{bansal2019beyond}. Experimental design in \cite{Zhang_2020} includes exposure to an AI that is comparable in performance to humans. Post-experimental observations highlighted that the error boundaries of the AI and the human were largely aligned (i.e., the humans and the AI were making the same types of mistakes), which limited the ability of AI confidence information to improve Human-AI decision-making outcomes. Earlier work~\cite{tan2018investigating} investigating similar questions in a recidivism prediction domain found that differences between algorithmic and human error boundaries were not sufficient to be leveraged in a way that could significantly improve hybrid decision-making. Recent work accounts for the differences in error types between people and AI and targets training ML models that are more complementary to human skills \cite{wilder2020learning}. Our work instead looks at the differences in human and AI error boundaries from the simple but widely used lens of true positive and negative rates and then draws conclusions for achieving complementarity in these dimensions. 

\subsubsection{Algorithmic Aversion and Appreciation} 

Gaube et al. ran a study where physicians and radiologists were given chest X-rays and diagnostic advice \cite{gaube2021ai}. All advice was generated by human experts, but for some participants, this advice was labeled as generated from an AI system. In this work, they found that radiologists rated the advice as lower quality when it came from an AI system, but physicians, who had less task expertise, did not. This work highlights the phenomena of algorithmic aversion and algorithmic appreciation for users of different expertise. 

Algorithmic aversion is the phenomenon where people tend to rely more on human advice than algorithmic advice, even when the algorithm proves to be more accurate than the humans \cite{dietvorst2015algorithm}. Prior research has found that factors such as algorithmic error and domain of judgment may cause algorithmic aversion. Dietvorst et al. finds that people are averse to algorithmic forecasts after seeing them make an error, even when they notice that the algorithm model outperforms human forecast \cite{dietvorst2015algorithm}. This is potentially due to human beliefs that people are more likely to be better than algorithms and that humans are able to be perfect, which leads to a desire for perfect forecasts \cite{dawes1979robust, einhorn1986accepting, highhouse2008stubborn}. Another aspect is the error rate, as humans are inclined to overestimate a machine's perceived error rate even if it is wrong only occasionally \cite{dzindolet2002perceived, hoff2015trust}. Consequently, people have less tolerance for errors caused by algorithmic systems than error caused by humans \cite{dietvorst2015algorithm}.

Another factor that potentially causes algorithmic aversion is the domain of judgment.
People want to understand where a recommendation is coming from, and find that they have more in common with human-based recommendations as opposed to algorithms \cite{dawes1979robust, yeomans2019making, prahl2017understanding}. This has led to development of algorithms that are programmed to act more like humans \cite{madhavan2007similarities}. 

There is also increasing work on potential strategies to overcome algorithmic aversion, such as the human-in-the-loop strategy. This strategy suggests that individuals are more likely to use and rely on algorithms when they have the opportunity to possibly correct and minimize errors in their decision-making judgement, even when the algorithmic-assistance is imperfect \cite{dietvorst2018overcoming}. In a parallel line of work, recent research has shown that individuals are not always averse to algorithms \cite{logg2019algorithm}. Algorithm appreciation, also known as automation bias, refers to when people rely on equivalent forecasts made by an algorithm more heavily than one made by a human \cite{logg2019algorithm}. Various factors that cause algorithmic appreciation outlined below.

Djikstra et al. found that individuals believe algorithmic systems represent a more objective and rational perspective than a human being \cite{dijkstra1999user, dijkstra1998persuasiveness}. Additionally, environmental influences, such as time-critical situations, can influence an individual's reliance on algorithmic advice over humans \cite{robinette2017effect}. Yeomans et al. found that individuals are more likely to rely on algorithmic decision aides when the algorithm is transparent \cite{yeomans2019making}. Moreover, additional information from other users, such as their prior experience with the model, also has a positive effect on an individual's adoption of algorithms, helping users reduce their hesitancy towards algorithms and better able to assess the reliability of a decision aid \cite{alexander2018trust, dietvorst2018overcoming}. 

Misuse of algorithms is described as the failure that results when individuals incorrectly rely on algorithms. This behavior is salient when an individual's trust surpasses the algorithm's real capabilities \cite{lee2004trust,buccinca2021trust,chiang2021you,bansal2019updates}. While algorithmic trust can lead to positive outcomes when the algorithmic aid provides the right advice, this misuse of algorithms can result in negative outcomes when the AI is incorrect.
\subsection{Stall Catchers}
Citizen science is a collaboration between scientists and volunteers to support advancement of scientific research. A citizen science project can involve one person or millions collaborating towards a common goal, and public involvement is typically in data collection, analysis, or reporting. Some long-term citizen science platforms and projects include eBird ~\cite{sullivan2014ebird}, Zooniverse~\cite{simpson2014zooniverse}, and iNaturalist~\cite{nugent2018inaturalist}.

\begin{figure}[t]
    \centering
    \includegraphics[width=7cm]{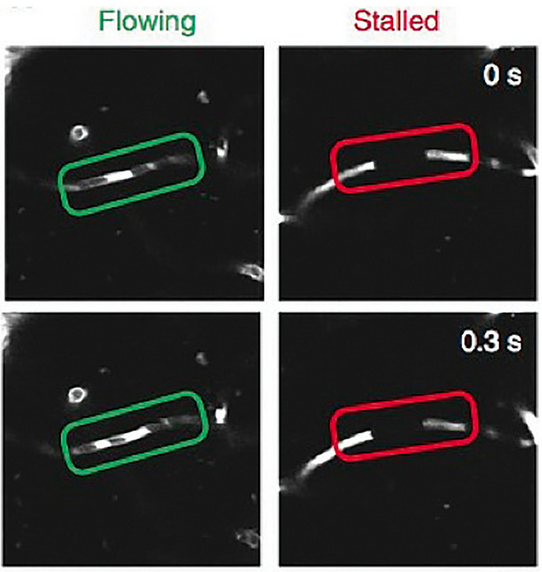}
    \caption{Side-by-side view of "flowing" vs "stalled" blood vessels from the Stall Catchers game.}
    \label{fig:stallcatchers}
\end{figure}

Stall Catchers is an online citizen science game created by the Human Computation Institute that crowdsources the analysis of Alzheimer's disease research data generated by Schaffer-Nishimura Laboratory at Cornell University's Biomedical Engineering department. Volunteers are tasked to watch and analyze video clips generated from \textit{in vivo} images of mouse brains in order to classify vessels as flowing or stalled (clogged blood vessels) as seen in Figure \ref{fig:stallcatchers}. In these videos, a single blood vessel segment is designated for analysis with a colorful boundary generated by a preprocessing algorithm. As users search for and "catch" stalls, they accumulate points. As users build up their scores, they advance in levels, compete for leaderboard spots, and may receive digital badges for their achievements in the game. The crowd-analyzed dataset reduces the number of blood vessel segments that lab-experts need to analyze for advancing Alzheimer's disease research by a factor of 20. Stall Catchers employs a so-called "wisdom of the crowd" algorithm to sensibly aggregate several judgements about the same vessel segment to produce a single, expert-like answer, which has been validated to be at least as accurate as that of a trained laboratory technician. Today, Stall Catchers has over 42,000 registered users who have collectively contributed over 12 million individual annotations, resulting in 1.4 million crowd-based labels. Several Alzheimer's research results enabled by Stall Catchers analysis have been published in top tier journals \cite{falkenhain_pilot_2020, michelucci2019people} and where possible, have listed "Stall Catchers players" as a co-author with a hyperlink to the list of usernames corresponding to Stall Catchers volunteers who specifically contributed to the results being published, in a rank-ordered list based on volume of relevant annotations (e.g., \cite{bracko_high_2020,falkenhain_pilot_2020}). Stall Catchers data has also been used to support a machine learning (ML) competition created in partnership with DrivenData and MathWorks which engaged over nine hundred participants and resulted in more than fifty machine learning models that can analyze Stall Catchers data~\cite{lipstein2020meet}. As with human Stall Catchers players, none of the resultant ML models achieved expert-like performance. These models did, however, exhibit a range of true positive and negative rates, which raises the possibility of Human-AI collaboration in the manner explored by research questions in the present study. 

In this work, we utilize a codeless experimentation toolkit developed by the Human Computation Institute, which executes an experimental design utilizing a sandbox version of Stall Catchers. This experimentation platform provides a proxy for problems relating to Human-AI interaction in medical decision-making. Several modifications were made to the experimental version of the Stall Catchers interface: 1) all gamification accoutrements (e.g., leaderboard, score, in-game chat, etc.) were stripped from the interface to reduce bias that might originate from point incentives, 2) a progress bar was added to the top of the interface to indicate the current stage and their progression through that stage (Figure \ref{fig:interface_withfeedback}), 3) the requirement and interface for indicating the location of an identified stall was removed, and 4) a small icon of a robot, with a hand pointing to either the "Flowing" or "Stalled" button, indicated the AI-Assistant's recommendation (Figure \ref{fig:interface_withAI}). We describe the details of these changes and setup in the following section.

\section{Study}
\label{sec:study}
\subsection{Participants} 
Recruitment of participants was done in compliance with the Human Computation Institute's IRB policies. We recruited 58 remote participants located in the USA, aged 18 years or older from the existing pool of players on the Stall Catchers citizen science platform~\cite{stallcatchers}. This recruiting method ensured that all players were familiar with the task. Participants had varying levels of expertise from novice to expert users and ranged in age from 18 to 82 years of age (mean = 49, sd = 17.2). No other user demographic data was collected on the Stall Catchers platform. Each participant completed the task in four stages, details of which are explained in Section~\ref{stages}. In each stage, participants watched videos of blood flowing through vessels of mice brains to identify if the vessels were flowing or clogged ("stalled"). Each participant was only allowed to participate in the task once to avoid any learning bias across different conditions. As an incentive for their voluntary participation, all participants received 5X Stall Catchers points for every correct answer, which were then added to their overall Stall Catchers player points. 


\subsection{Videos}
140 videos were sampled from a database with a validation set of 7,693 videos containing ground truth labels verified by expert players (i.e., players whose sensitivity was 1) not involved in the study. Each video contained cross-sectional views of blood flowing through vessels in mice brains (Figure~\ref{fig:interface_withfeedback}). Each video also had an associated difficulty metric (0,1) which was defined based on average response classification accuracy from historical Stall Catchers data.  

\subsection{Task Stages}~\label{stages}
The 140 videos were split across the four stages (20, 40, 40, 40) of the task, balanced by difficulty metric. Each stage contained equal proportions of stalled and flowing vessels. Details of each of the four stages are discussed below. 

\subsubsection{\textbf{Stage 1: No AI, With feedback}} 
\begin{figure}[ht]
    \centering
    \includegraphics[width=10cm]{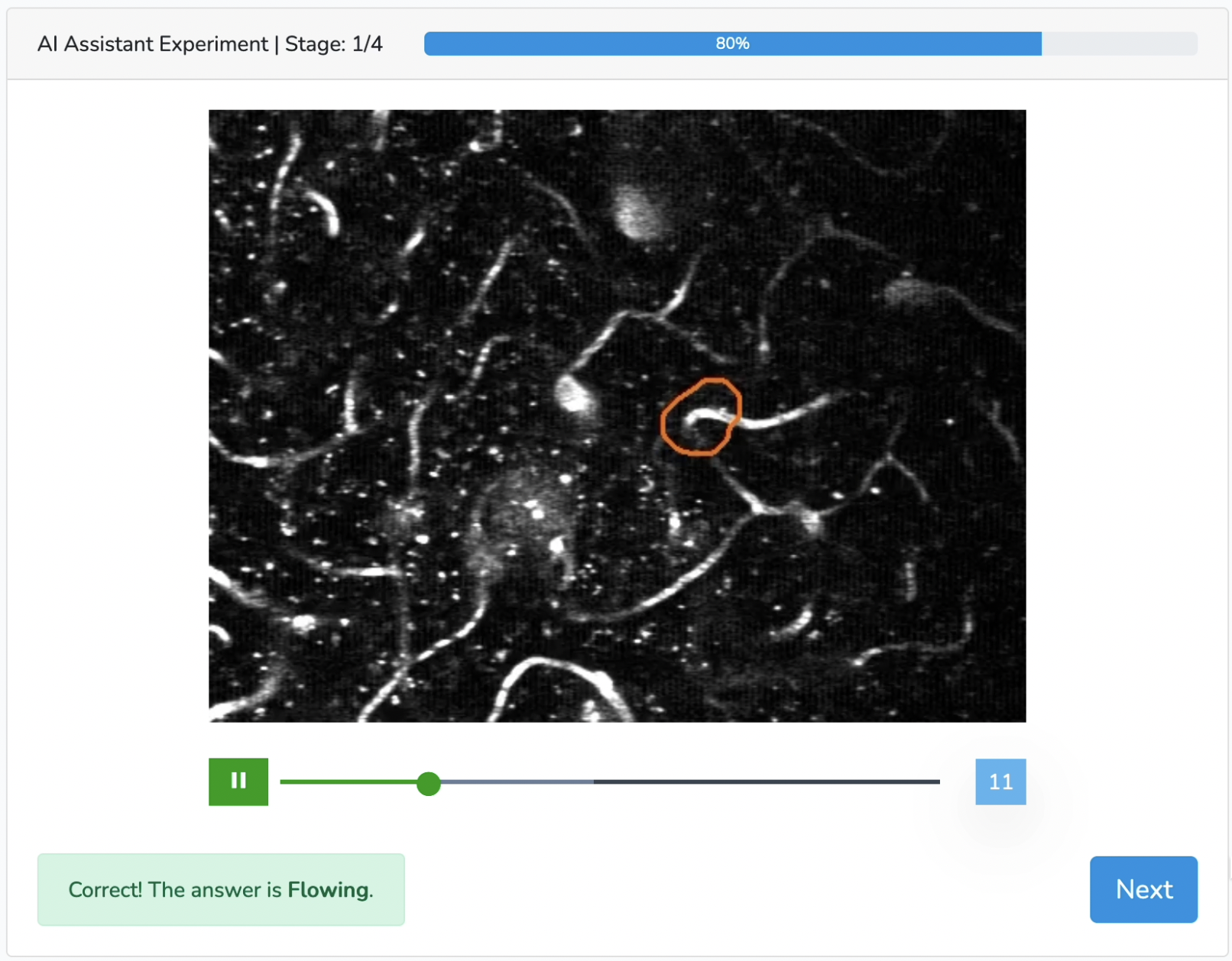}
    \caption{Stall Catcher interface showing feedback after a decision was made during Stage 1 and 3.}
    \label{fig:interface_withfeedback}
\end{figure}

In this stage, participants were shown 20 videos (10 flowing, 10 stalled) and asked to make a decision on each. Participants did not receive any form of assistance. After providing a response to a video, participants were shown the correct answer (Figure~\ref{fig:interface_withfeedback}). This stage served as a practice stage. Additionally, in the original Stall Catchers game, players are 4 times more likely to be shown a flowing video than a stalled video. Hence, this stage also aimed to override such prior biases that users may have had from playing the original game. 


\subsubsection{\textbf{Stage 2: No AI, No feedback}} In this stage, participants were shown 40 videos (20 flowing, 20 stalled) in randomized order and asked to make a decision alone.  Participants were not shown the correct answers after their selections. Responses from this stage were used to study the performance and skill level of participants when no AI assistance is present. 

\subsubsection{\textbf{Stage 3: With AI, With feedback}} 
\begin{figure}[t]
    \centering
    \includegraphics[width=10cm]{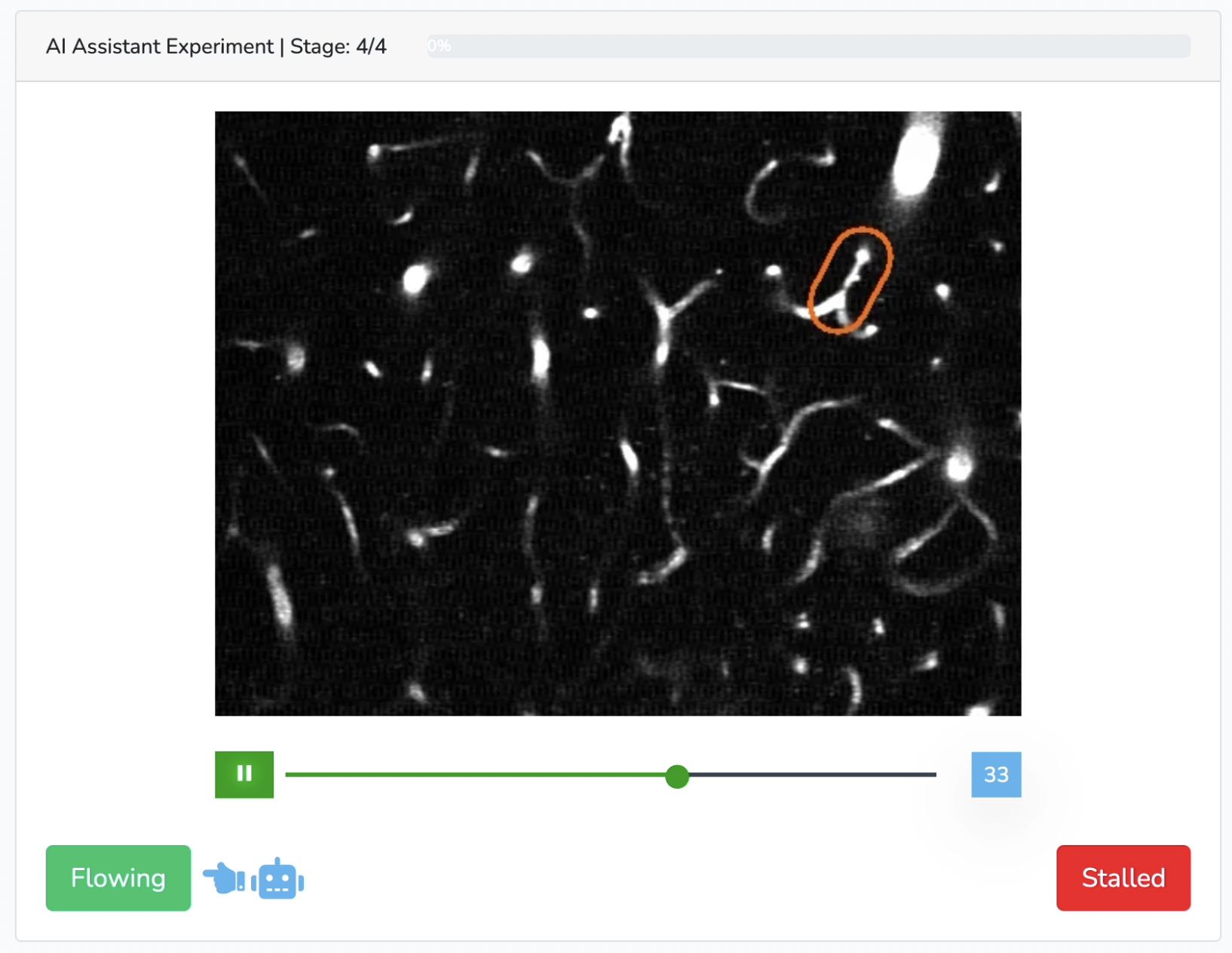}
    \caption{Stall Catcher interface with AI-Assistant suggestions, prior to the user making a selection, during Stage 3 and 4.}
    \label{fig:interface_withAI}
\end{figure}

In this stage, participants were shown 40 videos (20 flowing, 20 stalled) with randomized ordering alongside a recommendation from an AI assistant (Figure \ref{fig:interface_withAI}). Participants were required to make a final decision and could choose whether to incorporate the AI assistant's recommendation or ignore it. After participants indicated their decision, they were immediately given feedback on the correct answer. Stage 3 allowed for users to practice decision-making with the AI-Assistant and gauge the performance of the AI-Assistant. 

\subsubsection{\textbf{Stage 4: With AI, No feedback}} Stage 4 was the test stage where participants were shown 40 videos (20 flowing, 20 stalled) with randomized ordering alongside the AI-Assistant's recommendation and asked to make a decision. Participants were not shown the correct answers after their selections. Responses from this stage were used to assess the Human-AI team performance.

\subsection{Experimental Design}
A mixed factor design was used, with one between subjects variable (each participants was randomly assigned to different AI tuning conditions) and one within subjects variable (each participant was required to complete the task with and without AI assistance). The 58 unique participants were randomly assigned to one of three experimental conditions—Balanced condition, High TNR (True Negative Rate) condition, or High TPR (True Positive Rate) condition—while ensuring that each condition had similar numbers of participants. However, participants were not balanced across their prior expertise levels on the main Stall Catchers site. 

\subsubsection{Experimental Conditions}
The AI-Assistant in this study was simulated to enable specific control over the AI-Assistant's performance. Participants were randomly assigned to one of three experimental conditions varying how the AI assistant was simulated: Balanced, High TNR, High TPR. In all three conditions, the AI-Assistant's accuracy was kept constant at 75\% while the true positive and negative rates of the AI-Assistant changed in each condition. Table~\ref{AI_tuning_by_condition} summarizes the number of true negatives and positives for each condition along with accuracy, true positive rates, and true negative rates. We simulated the AI performance rather than tuning it for different true positive and negative rates so that we could exactly control these rates for experimentation across stages. While precise control in the AI tuning may not always be realistic or feasible in small samples, in this study, we aim at understanding the broad influence of different tuning factors. 

\begin{table}[h!]
 \centering
  \caption{AI-Tuning in Experimental Conditions.}
  \begin{tabular}{lp{0.5cm}c|cc|ccc}\toprule
  & \multicolumn{2}{c|}{AI Recommendation = Flowing} & \multicolumn{2}{c|}{AI Recommendation = Stalled} & Accuracy  & TNR & TPR  \\

            & TN & FN & FP & TP & $\frac{\mathrm{TN} + \mathrm{TP}}{\mathrm{TN}+\mathrm{FP} + \mathrm{TP}+\mathrm{FN}}$ & $\frac{\mathrm{TN}}{\mathrm{TN}+\mathrm{FP}}$ & $\frac{\mathrm{TP}}{\mathrm{TP}+\mathrm{FN}}$\\ \midrule 
   Balanced & 15 & 5 & 5 & 15 & 75\% & 75\% & 75\% \\
   High TNR & 19 & 9  & 1 & 11 & 75\% & 95\% & 55\% \\
   High TPR & 11 & 1  & 9 & 19 & 75\% & 55\% & 95\% \\
  \bottomrule
  \end{tabular}
  \label{AI_tuning_by_condition}
\end{table}

In the \textit{Balanced} condition, users interacted with an AI-Assistant that made an equal number of errors on each type of video (stalled or flowing). The AI-Assistant made five false positive errors, suggesting that a flowing video was stalled, and five false negative errors, suggesting that a stalled video was flowing, in each stage it appeared in. This resulted in the AI-Assistant having a TPR of 75\% and TNR of 75\%.

In the \textit{High True Negative Rate (TNR)} condition, the AI-Assistant was more likely to miss stalled videos. The AI-Assistant suggested that videos were flowing more frequently, resulting in a true negative rate of 95\% and true positive rate of 55\% for the AI-Assistant. 

In the \textit{High True Positive Rate (TPR)} condition, the AI-Assistant was more likely to miss flowing videos. The AI assistant suggested that videos were stalled more frequently, resulting in the AI-Assistant having a true negative rate of 55\% and true positive rate of 95\%.

As described in ~\ref{stages}, each participant completed the task in four stages, where responses from Stage 2 were used to determine participants' skill levels, and responses from Stage 4 were used to assess the Human-AI team performance. Thus, the set of videos in Stage 2 and Stage 4 were counterbalanced to help minimize any effects from task difficulty, i.e., half the participants received one set of videos in Stage 2, while the other half of the participants experienced that same set of videos in Stage 4.

\subsection{Data Collection and Procedures} At the start of the experiment, participants were shown an instruction page and a  participation consent form. After providing consent, each participant completed the video labeling task in each of the four stages in Section ~\ref{stages}. Participants were allowed to take multiple 2-min breaks, but were expected to complete the experiment within one sitting. Data collected included the participant's decision for each video shown, along with the AI-Assistant's recommendation (in Stages 3 and 4), and the ground truth for that video. Additional data recorded included the instant at which the user submitted their labels for each video (with one-second precision) and the response time (in seconds), i.e., the time from when video loaded until user clicked a response button.

After each stage, participants completed a short survey to assess their self-rated performance in that stage and to assess the perceived utility of the AI-Assistant in that stage (if present) on a scale from 1-5. These questions were adapted from the study by Kocielnik et al.~\cite{kocielnik2019will}. After Stage 3 and 4, participants were also asked to indicate what types of errors they thought the AI was making (more false positives, more false negatives, or equal). At the end of Stage 4, users answered eight questions related to their use, perceptions, and trust of the AI-Assistant. Users indicated their agreement with these questions on a 5-point Likert scale. 

A modification was made to the survey questions early in the study. As a result, some survey data is missing for the first 10 participants.

\section{Results}  
\label{sec:results}
\subsection{Overall Impact of the AI-Assistant on Performance}
\begin{mdframed}[style=MyFrame]
\textbf{Result 1: AI-Assistant recommendations significantly improved users' accuracy regardless of the AI-tuning.}
\end{mdframed}

Users' accuracy in each condition is shown in Table \ref{table:performance}. Accuracy was measured as the percentage of correct answers (\(\displaystyle \frac{\mathrm{TP}+\mathrm{TN}}{\mathrm{TN}+\mathrm{FP}+\mathrm{TP}+\mathrm{FN}}\)). Results from Stage 2 and Stage 4 (the two non-feedback conditions) were used to assess the impact of the AI-Assistant on users' accuracy for the Stall Catchers task. A mixed-model, repeated measures ANOVA was used with users' performance in Stage 2 and Stage 4 as the within subjects factor and tuning of the AI model (experimental condition) as the between subjects variable. A significant difference was found between Stage 2 and Stage 4, with users achieving higher accuracy alongside the AI-Assistant's recommendations in Stage 4 ($F_{1,55} = 24.61, p <.001, \eta^2 = .31, 1-\beta = 1.00$). Additionally, all three tunings of the AI had a positive impact on users' accuracy and there were no significant differences between the experimental conditions ($F_{2,55} = .61, p=.55, \eta^2 = .02, 1-\beta = .15$). 

We also examined users' performance in Stage 1 and Stage 3 to determine if the improvements could have been attributed to a learning effect. Paired samples t-tests revealed no significant differences between Stage 1 and Stage 2 ($t_{1,57} = .001, p=.87$) or Stage 3 and Stage 4 ($t_{1,57} = 1.86, p=.068$), which suggests that users performance did not significantly change between stages, except when the AI-Assistant recommendations were added.
\begin{table}[t]
 \centering
  \caption{Users' performance in terms of accuracy for Stage 1 (without AI + feedback), Stage 2 (without AI + no feedback), Stage 3 (with AI + feedback), and Stage 4 (with AI + no feedback). Users' accuracy increased significantly with the addition of AI recommendations (Stage 2 to Stage 4, $p<.001$), but there was no significant main effect of experimental condition (p=.562). There did not seem to be learning effects over time as there were no significant differences between Stage 1 and Stage 2 ($p=.979$), or between Stage 3 and Stage 4 (p=.054).}
  \label{table:performance}
  \begin{tabular}{lcccc}\toprule
            & \multicolumn{4}{c}{Accuracy} \\
            & Stage 1 &Stage 2  &  Stage 3 & Stage 4 \\ \midrule 
   Balanced &  $70\%$ & $66\%$  & $78\%$ & $72\%$     \\
   High TNR &  $61\%$ & $66\%$  & $72\%$ & $72\%$    \\
   High TPR &  $69\%$ & $68\%$  & $76\%$ & $77\%$    \\
  \bottomrule
  TOTAL &  $67\%$     & $66\%$  & $75\%$ &  $^*73\%$  \\
  \end{tabular}
\end{table}

\subsubsection{Baseline Performance Clusters}

\begin{figure}[ht]
    \centering
    \includegraphics[width=12cm]{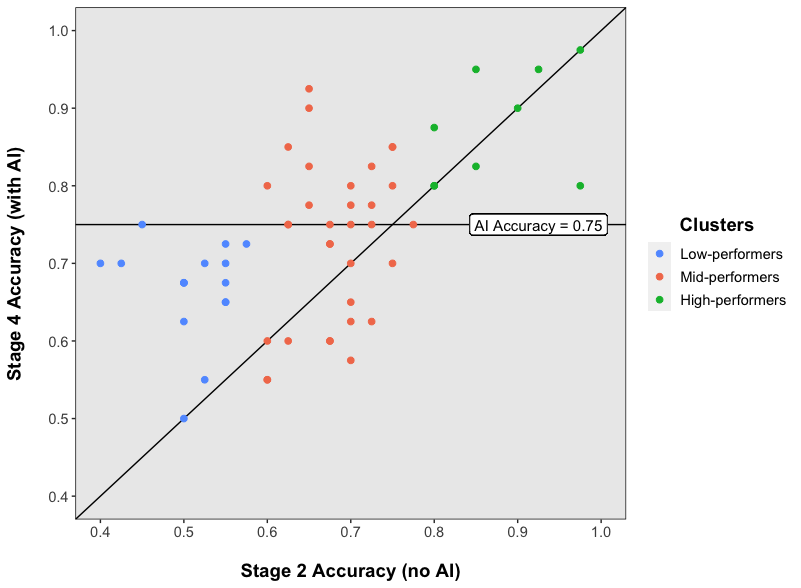}
    \caption{Accuracy of users in Stage 2 (without AI) and Stage 4 (with AI), clustered by users' accuracy in Stage 2 as a measure of human performance. Points above the diagonal line represent users who improved in accuracy, while points below the diagonal line represent users whose performance decreased. The horizontal line at 75\% accuracy represents the accuracy of the AI-Assistant. Data points above the horizontal line represent users who performed better than the AI-Assistant in Stage 4 and data points below the horizontal line represent users who performed worse than the AI-Assistant in Stage 4.}
    \label{fig:s2_s4_accuracy_all}
\end{figure}

A Pearson correlation coefficient was calculated to examine the performance gains made between Stage 2 (without AI) and Stage 4 (with AI). Across all users, there was a significant negative correlation ($r(57)=-.59, p<.001$) between Stage 2 baseline accuracy and gains made in Stage 4. Users with lower levels of accuracy made higher gains in performance when working with the AI-assistant compared to users with higher levels of accuracy. Figure \ref{fig:s2_s4_accuracy_all} illustrates the accuracy of participants in both stages. The diagonal line indicates the boundary at which participants would be equally accurate in both stages. Visually, improvements in Stage 4 result in data points above the diagonal line. Improvements above the accuracy of the AI-Assistant result in data points above the 75\% horizontal line.

Due to the interaction effect between baseline accuracy and improvement with AI, a two-step cluster analysis was used to group users based on their baseline accuracy in Stage 2. This resulted in three distinct performance clusters with a silhouette measure of cohesion of 0.7 (good), and cluster centers (means): Cluster 1 ($\bar{x}=51\%$), Cluster 2 ($\bar{x}=68\%$) and Cluster 3 ($\bar{x}=88\%$). 

One-sample t-tests were used to compare mean accuracy of each cluster in Stage 2 to the constant 75 percent accuracy of the AI-Assistant. The first cluster ($n=16$), termed the \emph{low-performer} group, had significantly lower accuracy than the AI-Assistant ($t_{15} = -19.62, p<.001$). The second cluster ($n=33$), termed the \emph{mid-performer} group, was closer to, but still significantly below the accuracy of the AI ($t_{32}= -7.96, p<.001$). The third cluster ($n=9$), termed the \emph{high-performer} group, had a significantly higher baseline accuracy than the AI assistant ($t_8 = 5.22, p<.001$). These resulting clusters were used for all subsequent analyses.
\clearpage

\subsection{Impact of AI-Assistant by Performance Cluster}
\begin{mdframed}[style=MyFrame]
\textbf{Result 2a: Regardless of AI-Assistant tuning, the accuracy of low-performers significantly increased, though not to the level of the AI-Assistant.}\newline \\
\textbf{Result 2b: Recommendations from the AI-Assistant had a mixed impact on mid-performers, with some users improving significantly and others getting worse. The High TPR condition was significantly better for mid-performers and resulted in more users achieving higher accuracy than the level of the AI-Assistant.} \newline \\
\textbf{Result 2c: High-performers maintained their high levels of accuracy, regardless of AI-Assistant presence or subsequent tuning.}
\end{mdframed}

\begin{figure}[ht]
    \centering
    \includegraphics[width=12cm]{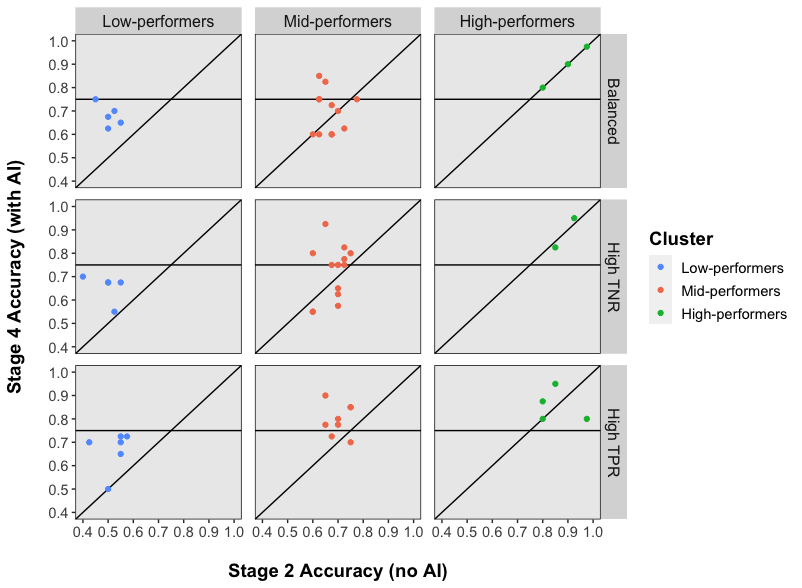}
    \caption{Accuracy of low-, mid-, and high-performers in Stage 2 and Stage 4 for each experimental condition.}    \label{fig:s2_s4_accuracy_bycluster}
\end{figure}

The effects of the AI-tuning were examined for each performance cluster (see Table \ref{table:performance_by_cluster}). Mixed repeated measures ANOVAs were conducted for each cluster.  Users' performance in Stage 2 and Stage 4 was the within subjects factor, and tuning of the AI model (experimental condition) was the between subjects variable. In addition, single-sample t-tests with a reference value equal to the accuracy of the AI-Assistant (75 percent) were used to analyze each cluster's Stage 4 accuracy relative to the accuracy of the AI-Assistant. Figure \ref{fig:s2_s4_accuracy_bycluster} shows users' accuracy results for Stage 2 and Stage 4, by performance cluster and experimental condition.\\

\begin{table}[t]
 \centering
  \caption{Users' performance in terms of accuracy for Stage 1 (without AI + feedback), Stage 2 (without AI + no feedback), Stage 3 (with AI + feedback), and Stage 4 (with AI + no feedback) by performance cluster. Low-performers significantly improved in accuracy from Stage 2 to Stage 4 ($p<.001$). Mid-performers also significantly improved in accuracy from Stage 2 to Stage 4 ($p<.01$). }
  \label{table:performance_by_cluster}
  \begin{tabular}{lcccc}\toprule
            & \multicolumn{4}{c}{Accuracy} \\
            & Stage 1 & Stage 2  &  Stage 3 & Stage 4 \\ \midrule
   Low-performers &$53\%$    & $51\%$  &  $70\%$ & $^*67\%$     \\
   Mid-performers &$68\%$     & $68\%$  & $74\%$ & $^*74\%$    \\
   High-performers &$83\%$        & $88\%$ & $89\%$ & $88\%$    \\
  \bottomrule
    \end{tabular}
\end{table}

\subsubsection{Low-Performers}
Recommendations from the AI-Assistant significantly improved low-performers accuracy ($F_{1,13} = 50.08, p<.001, \eta^2 = .79, 1-\beta = 1.00$). Of the 16 low-performers, 15 improved their accuracy, 1 stayed the same, and there were no low-performers whose accuracy degraded. Low-performers still performed significantly worse in Stage 4 than the AI-Assistant ($t_{15} = -5.15, p<.001$). No significant effect of experimental condition (AI-tuning) was found for the accuracy of low-performers ($F_{2,13} = 0.41, p<.674, \eta^2 = .06, 1-\beta = 0.10$).

\subsubsection{Mid-Performers}
\label{sec:mid_performer_results}
Recommendations from the AI-Assistant significantly improved mid-performers' accuracy ($F_{1,30} = 8.77, p=.006, \eta^2 = .23, 1-\beta = .82$), raising them to a comparable level of accuracy to the AI-Assistant in Stage 4 ($t_{32} = -1.14, p=.261$).  

Although there was a significant overall effect, improvements in accuracy were highly variable for mid-performers with some users improving and others falling below their baseline Stage 2 accuracy level. In Stage 4, 20 out of the 33 mid-performers improved their accuracy, two stayed the same, and 11 degraded in accuracy compared to their Stage 2 accuracy. Of the 20 users that improved in accuracy, 13 improved above the level of the AI-Assistant, five improved to the level of the AI-Assistant, and two improved but were still below the accuracy of the AI-Assistant.

A significant main effect was also found for experimental condition for mid-performers ($F_{2,30}=3.52, p=.043, \eta^2 = .19, 1-\beta = 0.61$). We examined each AI tuning condition separately using paired t-tests. In both the Balanced condition and the High TNR condition, overall mean accuracy rates increased when the users were given recommendations from the AI-Assistant, but this difference was not statistically significant given the high variance (($t_{11}=-1.087, p=.30$, and $t_{12}=-1.175, p=.26$, respectively). In the Balanced condition, mean accuracy rates rose from $66\%$ in Stage 2, to $70\%$ in Stage 4. In the High-TNR condition, mean accuracy rose from $68\%$ in Stage 2 to $72\%$ in Stage 4. However, the AI-Assistant recommendations did significantly improve mid-performers' accuracy in the High TPR condition ($70\%$ to $80\%$, $t_{7}=-3.188, p=.015$)

Examining mid-performer performance in Stage 4, we also see interesting trends across experimental conditions. Performance was mixed in the Balanced condition, where five users improved their accuracy from Stage 2 to Stage 4, two performed similarly, and five fell below their Stage 2 score. The High TNR condition was similar, where eight users improved in Stage 4 and five whose performance degraded. However, in the High TPR condition, all of the users (n=7) improved in accuracy, except one whose performance degraded. 

To understand what conditions maximized Human-AI team performance, we examined which mid-performers were able to increase their accuracy \emph{above} the level of the AI-Assistant. Of all mid-performers, $17\%$ (n=2) improved their accuracy above the level of the AI-Assistant in the Balanced condition, $38\%$ (n=5) improved above the level of the AI-Assistant in the High TNR condition, and $75\%$ (n=6) increased their accuracy above the level of the AI-Assistant in the High TPR condition. These results suggest that mid-performers saw the biggest gains in accuracy when working with an AI-Assistant that was tuned for High TPR (i.e., an AI-Assistant that erroneously suggests more stalls). Given these results, we conduct further analyses in Section \ref{sec:complementarity_results} to understand whether human baseline performance interacted with the AI tuning experimental condition in a complementary manner to influence the accuracy rates of mid-performers. 

\subsubsection{High-Performers}
The AI-Assistant did not significantly change high-performers' accuracy between Stage 2 and Stage 4 ($F_{1,6} = 0, p=1, \eta^2 = .0, 1-\beta = .05$). High-performers continued to perform at levels above the AI-Assistant's accuracy in Stage 4 ($\bar{x} = 88 \%, t_8 = 5.22, p<.001$). Experimental condition also did not significantly affect the accuracy of high performers ($F_{2,6} = .29, p=.76, \eta^2 = .09, 1-\beta = 0.08$). We would like to note that access to experts was limited so the number of high-performers in our study was low. This may have limited the statistical power to detect significant differences for this group of users.
\clearpage
\subsection{Complementarity: The Impact of Algorithmic Tuning}
\label{sec:complementarity_results}

\begin{mdframed}[style=MyFrame]
\textbf{Result 3: Users benefited most from the AI assistance when the AI was tuned to be more complementary to human expertise. Users only increased significantly in a performance measure when their initial baseline was below the AI-Assistant's level for that performance measure.}
\end{mdframed} 

As mid-performers benefited most from a High TPR assistant in Section \ref{sec:mid_performer_results},
we examined users' baseline true positive and true negative rates to better understand the inherent strengths of users and the types of errors they make, as well as the impact of tuning an AI-Assistant for one of these measures. The true positive rate is the percentage of stalled videos detected correctly (\(\displaystyle \frac{\mathrm{TP}}{\mathrm{TP}+\mathrm{FN}}\)). The true negative rate is the percentage of flowing videos detected correctly (\(\displaystyle \frac{\mathrm{TN}}{\mathrm{TN}+\mathrm{FP}}\)). Mean TPR and TNR scores within experimental conditions for Stage 2, Stage 3, and Stage 4 are shown in Table \ref{table:specificity}.

\begin{table}[t]
 \centering
  \caption{Users' performance in terms of true positive and true negative rates for Stage 2 (without AI) and Stage 4 (with AI). In the Balanced and High TPR conditions, users' TPR increased significantly from Stage 2 to Stage 4 (Balanced: $p<.001$, High TPR: $p<.004$). In the High TNR condition, users' TNR increased significantly from Stage 2 to Stage 4 ($p<.001$).}
  \label{table:specificity}
  \begin{tabular}{lcc|ccc}\toprule
            & \multicolumn{2}{c|}{True Negative Rate} & \multicolumn{2}{c}{True Positive Rate}\\
            & Stage 2  &   Stage 4 & Stage 2  &  Stage 4\\ \midrule 
   Balanced  & $75\%$    & $79\%$  & $57\%$    & $^*66\%$ \\
   High TNR & $75\%$  & $^*84\%$ & $56\%$    & $60\%$ \\
   High TPR & $77$\%    & $81\%$  & $59\%$   & $^*72\%$  \\
  \bottomrule
  TOTAL &  $75\%$  &  $^*81\%$ & $57\%$  & $^*66\%$  \\
  \end{tabular}
\end{table}

\subsubsection{Human Tuning}
\begin{figure}[ht]
    \centering
    \includegraphics[width=12cm]{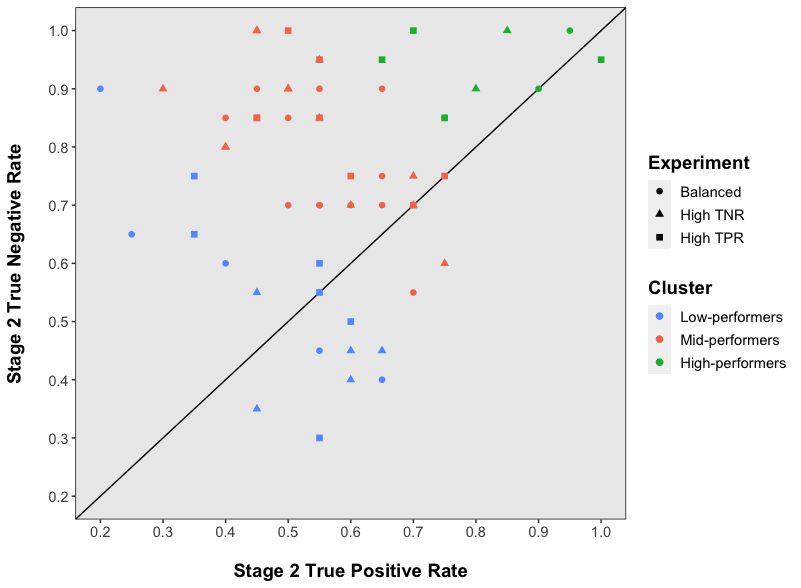}
    \caption{True positive and negative rates of all users in Stage 2. Points above the diagonal line represent users who were biased towards true negatives (better at detecting flowing) and points below the diagonal line represent users who were biased towards true positives (better at detecting stalled). }    \label{fig:s2_recall_specificity_clusters}
\end{figure}

We examined users' baseline true positive and negative rates from Stage 2 to determine if the users themselves were balanced in the errors they made, or if they were biased towards one of the metrics. Paired t-tests comparing users' scores in Stage 2 indicated that mid- and high-performers were better at detecting when vessels were flowing, leading them to have higher TNR compared to TPR ($mid: t_{32} = -7.33, p<.001; high: t_8 = -2.92, p=.019$). This can be seen in Figure~\ref{fig:s2_recall_specificity_clusters} where more of the data points from low and high-performers fall above the diagonal. Low-performers did not display a statistically significant bias towards either of the performance metrics ($low: t_{15} = -0.72, p=.485$).

\subsubsection{A Balanced AI-Assistant}

In the balanced condition, the AI-Assistant's true positive and negative rates were equal (75\% each) and therefore, made equal numbers of errors on stalled and flowing videos. Users in this condition had a baseline TNR of 75\% (not significantly different that the AI-Assistant, $t_{19} = -.069, p=.946$) and a baseline TPR of 57\% (significantly below the level of the AI, $t_{19} = -4.346, p<.001$). Paired t-tests were used to examine users' improvement in both scores from Stage 2 to Stage 4.  Users' true negative rates did not change significantly (79\%  $t_{19} = -1.182, p=.252$); however, users' true positive rates significantly increased (66\%, $t_{19} = -2.275, p=.035$), though still remained below the level of the AI-Assistant ($t_{19} = -3.383, p=.003$).

\subsubsection{A High TNR AI-Assistant}
In the High TNR condition, the AI-Assistant was tuned to have a TNR of $95\%$ and a TPR of $55\%$. The AI-Assistant was highly accurate when it predicted a stall, but was also more likely to suggest that a video was flowing and therefore, miss more stalls. The users in this condition had a baseline TNR of 75\% (significantly below the AI-Assistant, $t_{19} = -4.251, p<.001$) and a baseline TPR of 56\% (not significantly different from the AI-Assistant, $t_{19} = 0.383 p=.706$). Paired t-tests used to examine user improvement in Stage 4 revealed that users' TNR significantly increased ($84\%, t_{19} = -2.132, p=.047$), albeit still below the level of the AI-Assistant ($t_{19} = -3.222, p=.004$). Users' TPR in Stage 4 did not change significantly (60\%, $t_{19} = -0.892, p=.384$).

\subsubsection{A High TPR AI-Assistant}
In the High TPR condition, the AI-Assistant had a $95\%$ TPR and $55\%$ TNR. The AI-Assistant was highly accurate when it predicted a vessel was flowing, but was more likely to suggest that a video is stalled and therefore missed more flowing vessels. The users in this condition had a baseline TNR of $77\%$ (which was significantly better than the AI-Assistant, $t_{18} = 4.813, p<.001$) and a baseline TPR of $59\%$ (which was significantly below the level of the AI-Assistant, $t_{18} = -4.073, p<.001$). Paired t-tests used to examine user changes in TNR and TPR in Stage 4 revealed that users' TNR did not increase significantly ($\bar{x} = 81\%$, $t_{18} = -1.222, p=.238$). Users' TPR in Stage 4 did significantly increase ($72\%$, $t_{18} = -3.222, p=.004$) although was still below the level of the AI-Assistant ($t_{18} = -6.724, p<.001$).

\subsection{Agreement with AI-Assistant}
\begin{mdframed}[style=MyFrame]
\textbf{Result 4: Users were more likely to disagree with an incorrect recommendation (that flowing videos were stalled) in the High TPR condition. Additionally, users with higher baseline accuracy were more likely to agree with correct recommendations and disagree with incorrect recommendations.}
\end{mdframed}

We were also interested in studying user agreement with the AI-Assistant both when the AI is correct (i.e., both the human and the AI are correct) and when the AI is incorrect (i.e., both the human and the AI are incorrect). In the former case, agreement is beneficial to the Human-AI team but in the latter case, inappropriate agreement would lead to lower team performance. Non-parametric statistics were used to assess differences in agreement since the normality assumption was violated. Kruskal-Wallis Tests were used to examine differences in agreement between the experimental conditions with Mann-Whitney U tests for post-hoc pairwise comparisons of significant main effects.

\subsubsection{Agreement by Condition}
Table \ref{table_agreement_with_AI} shows the percentage of time users were in agreement with the AI-Assistant's recommendation for each experimental condition. No significant difference was found in terms of the overall level of user-AI agreement across experimental conditions ($\chi^2=3.213, p=.201$). 

In each experimental condition, the AI-Assistant was correct 75\% of the time and made incorrect recommendations 25\% of the time. Optimal Human-AI team performance occurs when users agree with the AI-Assistant when it is correct ($75\%$ of the time) and disagree with the AI-Assistant when it is incorrect ($25\%$ of the time). When the AI-Assistant was correct, the experimental condition had no significant impact on the level of agreement users had with those recommendations ($\chi^2=1.605, p=.448$). However, when the AI-Assistant was incorrect, experimental condition had a significant impact on user agreement with incorrect recommendations ($\chi^2=7.498, p=.024$). Post-hoc pairwise analyses revealed that users were less likely to agree with an incorrect recommendation (that "flowing" videos were "stalled") in the High TPR condition  compared to the High TNR condition ($Z=-2.605, p=.009$).

\begin{table}[h!]
 \centering
  \caption{User agreement with the AI-Assistant by experimental condition. The shaded columns indicate a beneficial action (user agreement when the AI was correct and user disagreement when the AI was incorrect). Significantly more users disagreed with an incorrect AI recommendation in the High TPR condition compared to the High TNR condition ($p<.017$).}
  \begin{tabular}{lc|cc|cc}\toprule
            &  Overall      & \multicolumn{2}{c|}{AI Correct} & \multicolumn{2}{c}{AI Incorrect} \\ 
            & Agreement &  Agreed & Disagreed & Agreed & Disagreed \\
            &  &  (ideal = 75\%) & (ideal = 0\%) & (ideal = 0\%) & (ideal = 25\%) \\
            \midrule 
   Balanced & 69\% & \cellcolor{lightgreen}58\% & 17\% & 11\% & \cellcolor{lightgreen}14\% \\
   High TNR & 73\% & \cellcolor{lightgreen}60\% & 15\% & 13\% & \cellcolor{lightgreen}12\% \\
   High TPR & 68\% & \cellcolor{lightgreen}60\% & 15\% & $^*8\%$ & \cellcolor{lightgreen}$^*17\%$ \\ 
  \bottomrule
  \end{tabular}
  \label{table_agreement_with_AI}
\end{table}

\subsubsection{Agreement by Performance Cluster}

Table \ref{table_agreement_with_AI_by_expertise} shows the percentage of time that users were in agreement with the AI-Assistant's recommendation by users' base-level performance. No significant difference was found in terms of the overall level of agreement across different levels of performance ($\chi^2=5.820, p=.054$). However, whether or not users accepted the correct AI recommendations or rejected the incorrect AI recommendations varied based on the user's performance cluster. 

When the AI-Assistant was both correct and incorrect, the base-level performance of the user had a significant impact on the level of agreement users had with each of these recommendations ($\chi^2=9.992, p=.007$, and $\chi^2=19.729, p<=.001$). Post-hoc pairwise analyses revealed that high-performers were significantly more likely to agree with correct recommendations from the AI-Assistant than either mid- or low-performers ($Z=-3.155, p=.002$ and $Z=-2.484, p=.013$, respectively).  Additionally, high-performers were significantly more likely to disagree with incorrect recommendations than either mid- or low-performers ($Z=-2.392, p=.016$ and $Z=-3.624, p<.001$, respectively). Finally, mid-performers were significantly more likely to disagree with incorrect recommendations than low-performers ($Z=-3.452, p=.001$).
\clearpage
\subsection{Users' Perception of the AI-Assistant}
\begin{table}[h!]
 \centering
  \caption{User agreement with the AI-Assistant by baseline performance. The shaded columns indicate a beneficial action (agreement when the AI was correct and disagreement when the AI was incorrect). Significantly more high-performers agreed with the AI when it was correct than either mid- or low-performers ($p<.017$). Significantly more high-performers disagreed with an incorrect AI recommendation than low- and mid-performers ($p<.017$). Significantly more mid-performers disagreed with an incorrect AI recommendation than low-performers ($p<.017$).}
  \begin{tabular}{lc|cc|cc}\toprule
             &  Overall      & \multicolumn{2}{c|}{AI Correct} & \multicolumn{2}{c}{AI Incorrect} \\ 
            & Agreement &  Agreed & Disagreed & Agreed & Disagreed \\
            &  &  (ideal = $75\%$) & (ideal = $0\%$) & (ideal = $0\%$) & (ideal = $25\%$) \\
            \midrule 
   Low-Performers & $74\%$ & \cellcolor{lightgreen}$58\%$ & $17\%$ & $16\%$ & \cellcolor{lightgreen}$9\%$ \\
   Mid-Performers & $67\%$ & \cellcolor{lightgreen}$58\%$ & $17\%$ & $^*10\%$ & \cellcolor{lightgreen}$^*15\%$ \\
   High-Performers & $73\%$ & \cellcolor{lightgreen}$^*68\%$ & $^*7\%$ & $^*5\%$ & \cellcolor{lightgreen}$^*20\%$ \\ 
  \bottomrule
  \end{tabular}
  \label{table_agreement_with_AI_by_expertise}
\end{table}

\begin{mdframed}[style=MyFrame]
\textbf{Result 5a: Users who perceived that the AI-Assistant's performance was better than their own were more likely to improve, while users who rated themselves higher than the AI-Assistant tended to degrade in their performance when provided with AI recommendations.}
\\ \\
\textbf{Result 5b: Users who improved with the AI-Assistant's recommendations rated the helpfulness and satisfaction of the AI-Assistant higher than those whose performance degraded. They also rated the AI-Assistant higher in terms of improving their ability to detect stalls.}
\end{mdframed}

Users were asked to rate their own performance as well as the AI-Assistant's performance on a five-point scale after Stage 2 (without AI), Stage 3 (with AI + feedback), and Stage 4 (with AI + no feedback) (see Table \ref{table:user_performance_ratings}).\footnote{Note: Eleven users did not fill out the survey so we report on data for the remaining 47 users.} Users' self-ratings of their own performance did not change significantly across the stages (Friedman test: $\chi^{2} = 4.59, p=.101$), however, their rating of the AI-Assistant's performance increased significantly from Stage 3 to Stage 4 (Wilcoxon: $Z = -5.26, p<.001$). These results were consistent across all of the experimental conditions.

\begin{figure}[ht]
    \centering
    \includegraphics[width=13cm]{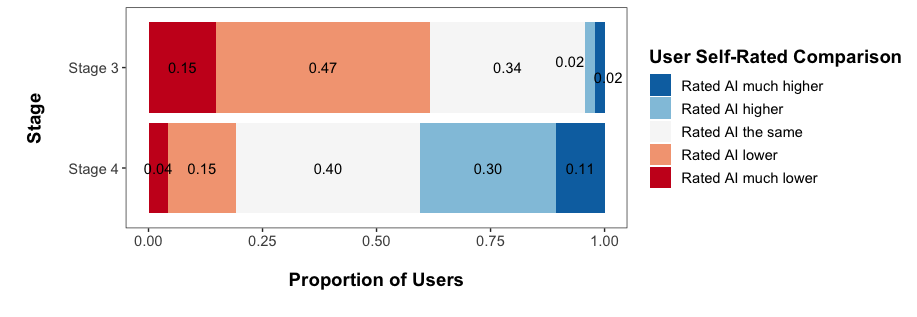}
    \caption{User' ratings of the AI-Assistant's performance compared to their own performance by stage.}
    \label{fig:s3_s4_AI_comparisons}
\end{figure}

We also examined users' relative ranking of their own performance in comparison to the AI-Assistant's performance in Stage 3 and Stage 4 (see Figure \ref{fig:s3_s4_AI_comparisons}). After Stage 3, when users were first introduced to the AI-Assistant and were also given feedback after every decision, users rated the AI-Assistant's performance significantly lower than their own performance (Wilcoxon: $Z = -4.325, p<.001$).  However, after further interaction with the AI-Assistant in Stage 4, users' rating of the AI performance increased significantly such that there was no significant difference between users' rating of themselves and the AI-Assistant at the end of Stage 4 ($Z = -1.766, p=.077$). 

\begin{figure}[t]
    \centering
    \includegraphics[width=13cm]{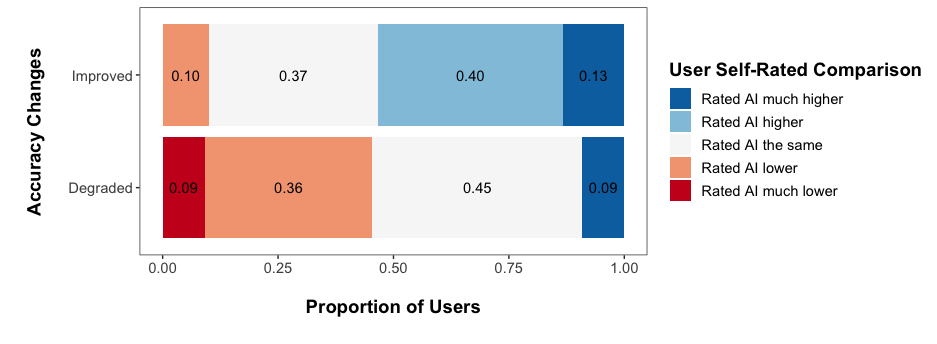}
    \caption{User comparisons of personal and AI performance based on accuracy improvements between Stage 2 and Stage 4}
    \label{fig:AI_comparison_all}
\end{figure}

The perceived performance level of the AI-Assistant may have impacted users' trust or reliance on the AI-Assistant's recommendations during Stage 4. Figure \ref{fig:AI_comparison_all} shows user ratings of the AI-Assistant for users whose performance improved from Stage 2 to Stage 4 versus users whose performance degraded from Stage 2 to Stage 4. Of the eleven users whose performance degraded, five rated the AI-Assistant performance similarly to themselves, five rated the AI-Assistant lower, and one rated the AI-Assistant higher. This resulted in a mean rating of their own performance ($\bar{x} = 3.55$) that was not significantly different than the mean rating of the AI-Assistant's performance ($\bar{x} = 3.09$) (Wilcoxon: $Z = -1.186, p=.236$). 
In contrast, for the thirty users who improved their accuracy from Stage 2 to Stage 4, eleven rated the AI-Assistant similarly to themselves, three rated the AI-Assistant lower, and sixteen rated the AI-Assistant higher. This resulted in a mean rating of their own performance ($\bar{x} = 2.87$) which was significantly lower than the mean rating of the AI-Assistant's performance ($\bar{x} = 3.43$) (Wilcoxon: $Z = -3.038, p=.002$).

\begin{table}[t]
 \centering
  \caption{Users' ratings of their own performance, and performance of the AI-Assistant for Stage 2 (no AI), Stage 3 (with AI + feedback), and Stage 4 (with AI no feedback) by experimental condition. Rating were on a 5-point scale with 1 being low and 5 being high. User ratings of the AI-Assistant increased significantly from Stage 3 to Stage 4 ($p<.001$), while ratings of their own performance did not change significantly (p=.101). }
  \label{table:user_performance_ratings}
  \begin{tabular}{lccc|cc}\toprule
            & \multicolumn{3}{c|}{User Self-Ratings} & \multicolumn{2}{c}{AI Ratings} \\
            & Stage 2 & Stage 3  &  Stage 4 & Stage 3 & Stage 4 \\ \midrule 
   Balanced & 2.90 & 3.53 & 3.12 & 2.47 & $^*3.35$     \\
   High TNR & 2.80 & 2.81 & 2.88 & 2.38 & $^*3.31$    \\
   High TPR & 2.94 & 3.21 & 3.21 & 2.50 & $^*3.29$  \\
   \bottomrule
   TOTAL & 2.88 & 3.19 & 3.06 & 2.45 & $^*3.32$ \\
  \bottomrule
  \end{tabular}
\end{table}

After Stage 4, users also rated the AI-Assistant's performance on a 5-point Likert agreement scale in terms of: 1) how helpful it was; 2) how satisfied they were with it; 3) how much it improved their ability to catch stalls; and 4) whether they would recommend the AI-Assistant to others. Responses to these questions were favorable, as $54\%$ reported that the AI-Assistant was helpful, $47\%$ were satisfied with it, $34\%$ felt that it improved their ability to find stalls, and $47\%$ would recommend it to others. A smaller percentage ($29\%$) indicated they they would use the AI-Assistant if it was available. Examining user ratings of the AI in terms of helpfulness, satisfaction, and improved ability, we again see a significant difference between users who improved in accuracy and those whose accuracy decreased. Users who improved in accuracy in Stage 4 rated the AI-Assistant higher in terms of being helpful (Mann-Whitney U: $Z = -2.27, p=.023$), being satisfied with it (Mann-Whitney U: $Z = -2.43, p=.015$), and feeling like it improved their ability to catch stalls (Mann-Whitney U: $Z = -2.917, p=.004)$.

\subsection{User Mental Models of AI-Assistant's Errors}
After Stage 4, when users were asked if they had a good understanding of how the Stall Catchers assistant decided whether a video contained stalled vessels or not, most users ($72\%$) said they disagreed or were unsure. Similarly, when asked if they knew what types of errors the AI-Assistant was making, most users ($72\%$) disagreed or were unsure. When asked whether the AI-Assistant was making more errors on "stalled" or "flowing" vessels after Stage 4, the majority of users ($62\%$) felt that the AI-Assistant made equal number of errors on both flowing and stalled vessels (Table  \ref{table_stage4_perceived_error_types}), regardless of the experimental condition. These results all suggest that users struggled with forming a mental model of the AI-Assistant and seemed unable to discern what types of errors the AI-Assistant was making. 

\begin{table}
\centering
 \caption{Users' Mental Model of AI-Assistant Error Types After Stage 4.}

  \begin{tabular}{lp{2.9cm}p{2.9cm}p{2.9cm}p{2.9cm}}\toprule
            Condition & more errors on 'stalled' vessels  & equal errors on flowing and stalled vessels &  more errors on 'flowing' vessels & makes no mistakes at all   \\ \midrule 
   Balanced & 3  & 12 & 4  & 0 \\
   High TNR & 4  & 11 & 2 &  0  \\
   High TPR & 1 & 10 & 5  &  0\\ \midrule
   TOTAL & 8 & 33 & 11  & 0 \\
  \bottomrule
  \end{tabular}
    \label{table_stage4_perceived_error_types}
\end{table}

\subsection{Self-Reported Use of AI}
After Stage 4, users were asked "How did you use the Stall Catchers assistant's recommendations (if at all)?" Although the question was optional, all participants responded. Two researchers coded all of the comments, using a bottom-up open-coding approach ($85\%$ inter-rater reliability) followed by Axial coding to cluster related comments.  Four key categories were identified: 1) follow recommendations if unsure, 2) take a second look / verify, 3) ignore, and 4) check stalled videos. The remaining comments were labeled "other." 

\subsubsection{Follow recommendations if unsure}
The most common way people used the AI-Assistant was when they were unsure with their decision. Nineteen users indicated that they consulted or followed the AI-Assistant in instances where they were unsure of what they viewed in the blood vessel video by themselves. For example, User 120 (high-performer, High TPR) responded ``\emph{I used the assistant to help me make a choice when I wasn't able to decide which appropriate answer to choose}," and User 158 (mid-performer, High TPR) said ``\emph{I only used the assistant if I couldn't decide if there was a stall or not, or if the image was too grainy to see one way or another.}" Users incorporating this strategy would likely benefit from a complementary AI where the AI is tuned to have higher accuracy in areas where users are weaker. There is support for this hypothesis in our data through examination of the eleven mid-performers who used this strategy. Five were in the non-complimentary Balanced condition, and four of these users performed worse in Stage 4. Two users were in the High TNR condition and both improved (one of whom improved above the level of the AI-Assistant). The final four users were in the High TPR condition, which was the most complimentary condition based on initial human tuning, and all four users significantly increased their accuracy above the level of the AI-Assistant.

\subsubsection{Take a second look / verify}
Ten users indicated that they would use the AI-recommendation to verify their decisions or re-visit the video if the AI-Assistant recommendation disagreed with their initial assessment. User 166 (low-performer, High TNR) responded ``\emph{I tried to ignore them and make my own choices and then look again if I disagreed}."  User 182 (low-performer, High TPR) wrote ``\emph{If it disagreed with my findings I would re-evaluate the movie. Sometimes it changed my mind and sometimes it didn't.}" This approach would also benefit from a complementary AI. If users and the AI both tend to make mistakes in the same areas, teams would not improve in performance after users took a second look in areas where they were weak. A complementary AI could be exploited by tuning the AI to prompt users to scrutinize or put more cognitive effort into their areas of weakness. 

\subsubsection{Ignore}
Six users reported that they ignored the AI-Assistant's recommendation, all of which were mid-performers, and four of these users' accuracy went down in Stage 4. User 169 (mid-performer, High TNR, accuracy dropped from 60\% to 55\%) shared ``\emph{After several movies I stopped paying attention to assistant's opinion.}".  User 170 (mid-performer, High TPR, accuracy dropped from 75\% to 70\%) also disregarded the AI-Assistant and said ``\emph{I didn't look at the recommendation until after I viewed the movie and decided for myself. I didn't really rely on the assistant.}" These results indicate that mid-performers struggled with whether or not to use the AI-Assistant, and those that chose to ignore it often degraded in performance.

\subsubsection{Check stalled videos}
Five users indicated that they mostly acted on information from the AI-Assistant when it recommended a stall. User 110 (mid-performer, High TNR) wrote ``\emph{If the assistant recommended a stall, I usually would go back over the movie a few more times to try and catch it, just to figure out what it saw to flag it as a stall. Often, this helped me catch a stall I missed, simply by suggesting it was a stall after all.}" User 107 (mid-performer, High TNR) wrote ``\emph{Only if I felt I could confirm them with my own eyes. Each time the assistant caught a stall that I didn't, I went back to find it}". Feedback from these users reinforces the utility of High TPR tuning (which is complementary to the users), as it suggests more stalls.

\section{Discussion}
\label{sec:discussion}
Consistent with findings from prior work \cite{kamar2016directions, wang2016deep, wang2021deep, steiner2018impact, gaur2016effects}, providing users with recommendations from an AI-Assistant significantly improved team performance on the Stall Catchers task. However, various experimental factors in our study influenced the extent of these gains such that some users had marginal improvements, some saw more significant improvements, and others degraded in performance. Although any improvements in performance are beneficial, performance gains that fall short of the level of the AI are common in prior work~\cite{lai2019human,bansal2021does,feng2019can,lundberg2018explainable}. Our goal was to understand how we can maximize Human-AI team performance such that it is significantly better than the level of the AI or the human alone. In this section, we reflect on the study results to provide insights and guidance on ways to maximize Human-AI team performance.

\subsection{Baseline Expertise} 
 The largest determining factor for superior Human-AI team performance was the base level expertise of the users. Benefits obtained from use of the AI-Assistant systematically varied with users' baseline expertise on the task. Individuals with low expertise significantly improved their performance by working with an AI-Assistant, though never to accuracy levels at or above the AI. In these cases, AI recommendations helped, but users lacked the knowledge or understanding of when they should accept or reject an AI recommendation. Studies that primarily recruit novice users (such as Mechanical Turk studies) likely experience similar issues. Conversely, the performance of individuals with high expertise was not impacted by the AI-Assistant, and these users continued to perform at accuracy levels well above the AI-Assistant. For individuals with mid-level performance the results were mixed, some improved above the level of the AI, and others performed worse. Understanding the performance gains and losses for this class of users provides important insights on how to effectively support Human-AI collaboration.  

\subsection{Tuning for Complementarity}
A second determining factor of success was related to tuning the AI algorithm for complementarity. Too often, the focus for an AI model is overall algorithmic performance or accuracy. As our work shows, even in situations when the overall accuracy level remains constant, different tunings of an algorithm can produce significantly different results. Differential tuning for false positives or false negatives reinforced or complemented users' strengths and weak spots. For example, in our study, mid-performing users benefited significantly when the algorithm was tuned for a high TPR as it complemented users' natural bias towards a high TNR. As a result, the High TPR condition produced more consistent improvements and resulted in more users achieving accuracy results above the level of the AI. 

The relative performance of the AI compared to the users also influenced gains obtained from working with the AI. Across all our experimental conditions, significant increases in a specific performance measure only occurred when the users' score started below the AI-assistant's tuning for that measure. For example, if users had a lower TPR, then an AI tuned for a high TPR would improve users' TPR as well as their overall accuracy. This finding suggests that tuning an algorithm to complement a user's weakness can improve user performance on that measure and increase overall Human-AI team performance.  

At the same time, tuning for complementarity must not sacrifice trust from the user in an AI tool. Lu \& Yin find that human reliance on a model is significantly influenced by their agreement with an AI during task instances when they have high confidence in their own decision \cite{lu2021human}. Thus, if an AI tuned to complement a user's weaknesses makes too many incorrect recommendations in the area of a user's strengths, human trust and reliance on the AI may decrease. As a result, there is likely a balance to be struck between complementary tuning and trust, which is discussed in the next section. 

Another important consideration related to algorithmic tuning includes users' strengths in identifying or overriding positive or negative cases. Optimal Human-AI team performance comes from users being able to accept instances when the AI is correct, and override instances when the AI is incorrect. By examining when users aligned with and ignored AI recommendations, we found that users in each experimental tuning condition were equally likely to agree with the AI when it was correct. However, when the AI was incorrect, users in the High TPR condition were more likely to override AI recommendations, likely because of an inherent strength in detecting flowing vessels. 

The Selective Accessibility (SA) model—a social cognition concept that attempts to explain the influence of anchor values and comparative judgements during human-decision making—might explain why our study found that the High TPR condition increased user performance ~\cite{mussweiler1999comparing}. The SA model posits that when users are given an anchor value during a task (in this case, the AI recommendation), users engage in \textit{hypothesis-consistent testing}, actively retrieving knowledge and their personal framework related to the judgement they are being asked to make. Given incorrect stall recommendations in the High TPR condition, users might have activated hypotheses related to whether the video was stalled, rather than using their standard decision-making framework to decide if a video is stalled or flowing. Since users were better at detecting flowing videos, it is possible that users had more robust evidence for factors that characterized a certain video. This prior may give them more success in rejecting stalled recommendations at the correct instances compared to flowing recommendations. If users had a stronger hypothesis or evidence requirements for one type of video, the SA model could explain why participants struggled to disagree with incorrect flowing AI suggestions in the High TPR condition. The SA model and the idea of hypothesis-consistent testing support our findings that complementary AI systems can enhance performance by nudging characteristics of a video to be more salient, accessible, or relevant to the user during decision-making.

Klayman \& Ha also find that the most critical way to test hypotheses is a "positive test strategy," where participants examine cases with the target characteristic and retrieve knowledge from their memory related to the assumption provided by an anchor \cite{klayman1987confirmation}. In our study, this strategy would involve users examining videos where the AI suggested a stall, and evaluating the video with their past memories and expectations. This finding is consistent with users who chose to focus their attention on positive AI suggestions and qualitatively reported their AI-usage workflow as "checked stalled videos." 


\subsection{Perceived Performance, Trust, Helpfulness, Satisfaction}
The third determining factor of superior performance for users in our study was the perceived performance of the AI-Assistant relative to their own performance. As shown in prior work \cite{lee2004trust,buccinca2021trust,Zhang_2020}, trust is a critical factor related to Human-AI team performance; however, its impacts are not well understood and can be difficult to measure reliably. For example, when a human agrees with the AI, it is difficult to distinguish whether they agreed because the human trusted the AI or because they independently solved the task without AI help and happened to agree. Many studies rely either on proxy measurements of trust (e.g., agreement, time of completion, usage, stickiness) or self-reported scores (e.g., satisfaction, perceived accuracy, recommending the assistant to others). We examined users' self-ratings of their own performance and performance of the AI to gauge users' perception of, and possibly trust in the AI-Assistant. 

Regardless of the absolute ranking, users who rated the AI performance above their own were more likely to benefit significantly from the AI recommendations, while users who rated the AI-Assistant and their own performance similarly tended to degrade in performance. It is possible that users who felt that the AI was performing better than themselves inherently trusted the AI-Assistant more, resulting in more adherence to the recommendations and superior performance. In addition, users' ratings of the helpfulness of the AI-Assistant and satisfaction with the AI-Assistant were also found to be positive indicators of success. 

This result is also supported by the Technology Acceptance Model (TAM), which suggests that when users are presented with new technology, factors including perceived usefulness (PU) and perceived ease-of-use (PEOU) influence their decision about when and how they use the new technology \cite{venkatesh2000theoretical}. Ranking the AI-Assistant high in terms of performance, helpfulness, and satisfaction all suggest that users perceived the AI-Assistant to be useful, which would contribute to their acceptance of the AI-Assistant and its recommendations.

\subsection{Mental Model Formation}
Mental models can help users form trust and accept recommendations from an AI \cite{Zhang_2020, bansal2019updates}; however, our findings suggest that users were unable to create an accurate mental model of the AI-Assistant's recommendations for the Stall Catchers task. Most users believed that the AI-Assistant made equal errors on "stalled" and "flowing" videos regardless of experimental condition. This suggests that the complementary tuning of the AI-Assistant may have been less effective in supporting mental model creation. 

Our study intentionally did not present any information to users about how the algorithm was tuned or its overall accuracy so that we would be able to observe the presented effects without biasing their perception. However, for real-world deployments, when reliable tuning scores are available, presenting them to users may support creation of a mental model about the AI. Previous work has emphasized the importance of providing such information in order to set expectations on performance \cite{amershi2019guidelines, kocielnik2019will, yin2019understanding}. However, further innovation is needed to detail these summaries beyond aggregate scores of performance (e.g., aggregate accuracy, TPR, TNR) potentially with interactive and on-demand visualizations that explain to users when and how the AI fails and what it is able to do for them.

An accurate mental model alone will not always lead to better Human-AI team performance. Even when a human understands when the AI is not reliable, they still need to solve the task accurately. This problem is even more relevant for multi-class predictions or content generation where the domain space of the solution is of a higher dimension. Again, other forms of assistance may be needed in such cases that can guide the team towards more detailed and informed reasoning (e.g., similar historical examples, outlier analysis, data exploration capabilities etc.). 

\subsection{Role of the AI-Assistant}
While the TAM suggests that perceived usefulness and ease-of-use can predict acceptance/usage of technology, it is also important to examine how users are working with and perceiving the AI system. Our data indicated that users incorporated a variety of strategies when working with an AI-Assistant. These self-reported strategies may have implications for user performance with AI, as prior work has shown performance differences when users view the AI system as a second reader versus an arbitrator  \cite{wang2021deep}. In our study, some users provided feedback that they followed the AI if they were unsure, hinting at its role as an arbitrator. Meanwhile, other users indicated that they would double check the video if the AI disagreed with them, hinting at the possibility that they used the AI as a second reader in the task. Two other users indicated that confidence in their decisions increased when the AI agreed with them. 

We also observed users who blindly followed the AI recommendations, suggesting overreliance on the AI. Algorithmic appreciation is generally correlated with overreliance, but it is also important to understand why users are displaying a certain behavior. For example, one user indicated that they followed the AI "out of laziness," which has more to do with their own effort as opposed to appreciation for the AI. We also had many users who indicated that they ignored or did not rely on the AI-Assistance, which could be a product of general algorithmic aversion or a result of the AI's individual performance. In either of these cases, methods to foster trust or increase reliance on the AI in complementary instances might be dependent on how or why the user is using the AI in a certain way. 

Since users in our study were not instructed to use the AI in any specific manner, these results indicate yet another factor that can influence Human-AI team performance. Users can have the same expertise level and mental models and end up with completely different results based on how they work with the AI, such as a colleague, an assistant, a tool, a distractor, etc. More work is needed to better understand the impact of these different strategies, which may also change as users become more comfortable working with AI systems.   

\subsection{Tuning for Tasks/Scenarios or Users}
Often, the ideal tuning for an AI model will depend on the task or scenario.  For example, in a criminal justice situation, the model may be tuned to minimize false positives so innocent people don't get put in jail, while in a medical context, a model may be tuned to minimize false negatives so a cancer diagnosis is not missed.  However, as we have seen from this study, if both the users and the AI have the same tuning, performance gains from the Human-AI team may be stifled. In our study, an AI tuning that complemented users resulted in the largest performance gains. Additionally, sacrificing accuracy on a measure that users had strengths in, did not cause a degradation in that measure.

There are several potential directions for improvement inspired by these findings differentiating AI impact based on the algorithmic tuning. First, these findings can inform model selection prior to deployment. Furthermore, since different users may have different levels of expertise and inherent biases, further benefits can be achieved by personalizing the AI tuning based on the human preference and expertise to deploy algorithms that are most complementary to the user. 

One potential challenge in tuning for user complementarity is that AI models are often trained via human annotations, and therefore, approximate user behaviors in aggregate. While recent work has aimed at training more complementary models~\cite{wilder2020learning}, often there may not exist enough complementarity in the available labels themselves to train models with these properties. Therefore, other forms of assistance to human experts may be needed such as providing additional information instead of offering a recommendation. Examples of additional information may include digested summaries of data distributions or presenting historical examples. 

Tuning AI systems for complementarity with human users may prove to be more complicated for tasks that extend beyond multi-choice classification, including text generation or information retrieval. As tasks become more complex, factors such as efficiency, productivity, and cognitive effort may influence optimal human-AI team performance. For example, in the context of text generation, there may be trade-offs in satisfaction and productivity between providing a long output or translation that the user can edit and a shorter, more precise output that requires more user input. Given the various ways in which users can incorporate AI recommendations into their decision making, the success of an AI tool for a more complex task may rest on its ability to complement an individual's decision making framework or workflow. Future studies should explore the role of human expertise and algorithmic tuning as it relates to more complex tasks, such as retrieval, ranking, translation, and image/text generation. 

\subsection{Limitations} 
While the results from this study provide important insights related to Human-AI decision making, there are several limitations which may have influenced our findings. First, given that we wanted to examine a complex, real-world task that users had experience with, the available population of users was limited. This resulted in smaller sample sizes, particularly for high performing users, and an unequal distribution of participant expertise across experimental conditions. This potentially constrains some of our conclusions since the overall statistical differences observed may have been influenced by the larger sample size in the mid- and low-performing clusters. We are also missing some survey data from the first eleven participants due to minor changes introduced in the survey after deployment.

Since recommendations from the AI-Assistant were shown at the beginning of each task, it was difficult to fully understand how much the AI-Assistant's recommendations influenced users. It is possible that in many instances where the user "agreed" or aligned with the AI, the user was going to select the response regardless of the AI's recommendation. At the same time, fully decoupling anchoring bias effects from intentional agreement \cite{rastogi2020deciding} would require experimentation with other types of workflows where the human takes a temporary decision first and then has a chance to revise it after getting assistance from the AI. The challenge still remains valid even for such workflows, however, given that humans could still anchor to second opinions, which calls for further analytical methods encouraging self-reflection and explicit argumentation on why the decision maker may or may not agree with the AI.

The presentation of the task in the citizen science Stall Catchers platform frames the task as "finding stalled vessels." This framing may have biased users towards focusing on stalled vessels, and also focusing on stalled recommendations from the AI-Assistant. There is some evidence for this in the qualitative survey data, as some users indicated that a stalled recommendation from the AI-Assistant made them take a second look to see if a stall was actually there. We did not receive reciprocal comments regarding users trying to verify flowing recommendations from the AI-Assistant.

\section{Conclusion and Future Work}
\label{sec:conclusion}
This work examined the impact of algorithmic tuning and human expertise in Human-AI team performance, as well as the influence of user perceptions of AI. Our study centered around a real-world citizen science platform, Stall Catchers, to understand the impact of assisting users in a complex decision-making task. Our results highlight how degrees of human expertise can significantly impact the potential value of AI assistance, as benefits from AI assistance are highly variable for users whose individual performance is similar to that of the AI. Furthermore, the work showed that there are opportunities to boost team performance by deploying models that are complementary to human capabilities and fostering Human-AI interactions that help users develop appropriate mental models, as well as trust and confidence in AI systems.

Informed by these findings, we envision several directions for future work. The differential impact of human expertise and algorithmic tuning suggests that different types of AI could be beneficial for different users. Personalization techniques have been extensively studied in the context of content or item recommendation but not sufficiently for customizing AI partners to users. Such techniques will require learning and updating user models over time and adjusting AI deployments to those changes. While historical data for users may be initially sparse, possible avenues can be explored to apply these approaches to groups of individuals based on their expertise, similar to the clusters of users that we explored in this study. Considering the challenges of assisting human experts posed in our findings, personalizations of AI assistants may even need to go beyond measures of true positive and negative rates—instead considering more fine-grained definitions of error boundaries to build models whose main goal is to provide assistance for problems and examples that are most difficult to experts based on input characteristics. For such an audience, productivity is another aspect of improvement. Even if the human remains equally accurate with AI assistance, assistance that helps users make accurate decisions faster can still be beneficial for particular use cases. Overall, we hope that these findings inspire efforts to improve AI systems with human-centered considerations in ways that best augment and complement users.

 \begin{acks}
This study would not have been possible without the contributions of Stall Catchers users. The authors would like to thank Ece Kamar for her assistance with this project and the anonymous reviewers for their useful feedback.  
\end{acks}

\bibliographystyle{ACM-Reference-Format}
\bibliography{base}
\end{document}